\documentclass[journal=nalefd,manuscript=article]{achemso}

\usepackage{amsmath}
\usepackage{amssymb}
\usepackage{bm}

\usepackage{graphicx}

\title{Room-temperature gating of molecular junctions using few-layer graphene nanogap electrodes}

\author{Ferry~Prins$^\S$}
\email{f.prins@tudelft.nl}
\author{Amelia Barreiro$^\S$}
\email{a.barreiromegino@tudelft.nl}
\author{Justus~W.~Ruitenberg}
\author{Johannes~S.~Seldenthuis}
\affiliation{Kavli Institute of Nanoscience, Delft University of Technology, Lorentzweg 1, 2628 CJ Delft, The Netherlands}
\author{N\'{u}ria~Aliaga-Alcalde}
\affiliation{ICREA, Instituci\'{o} Catalana de Recerca i Estudis Avan\c{c}ats, Departament de Qu\'{i}mica Inorg\`{a}nica, Universitat de Barcelona, Mart\'{i} i Frances 1-11, 08028 Barcelona, Spain}
\author{Lieven~M.~K.~Vandersypen}
\author{Herre~S.~J.~van~der~Zant}
\affiliation{Kavli Institute of Nanoscience, Delft University of Technology, Lorentzweg 1, 2628 CJ Delft, The Netherlands}

\footnotetext{$^\S$ these authors contributed equally}

\begin{document}

\begin{abstract}

We report on a method to fabricate and measure gateable molecular junctions which are stable at room temperature. The devices are made by depositing molecules inside a few-layer graphene nanogap, formed by feedback controlled electroburning. The gaps have separations on the order of 1-2 nm as estimated from a Simmons model for tunneling. The molecular junctions display gateable IV-characteristics at room temperature.

\end{abstract}


\pagebreak

Molecular electronics has been the subject of extensive research over the past decade\cite{moth-poulsen2009, song2011}, motivated by the appealing concept that molecules can be used as ultimate downscaled functional units in electronic circuits performing a variety of functions, including rectifiers,\cite{aviram1974, metzger1997} switches, \cite{blum2005} transistors,\cite{park2000, kubatkin2003, yu2004} or sensors.\cite{guo2006} To date, three-terminal experiments have mostly been carried out at low temperatures, whereas for applications room-temperature operation is desirable. Device-stability is a first requirement for this but at the same time remains one of the great challenges in this field. Gold, the preferred electrode material because of its noble character, has such high atomic mobility that at room temperature the nanoelectrodes are unstable.\cite{Prins2009} Recently it was shown that electrodes made from the more stable Pt can be used to overcome this issue,\cite{Prins2009, Prins2011} although gated transport at room temperature has not been demonstrated yet.

An alternative strategy for the fabrication of stable electrodes with nanometer separation is the use of (sp$^2$-)carbon-based materials. The covalent bond-structure gives stability up to high temperatures, far beyond room temperature. Another advantage is that it allows for a large variety of possibilities to anchor diverse molecules as compared to metallic electrodes. While with the latter thiol and amine linkage is widely used \cite{song2011}, the carbon-based materials can not only be functionalized covalently through organic chemistry techniques\cite{guo2006} but also via $\pi$-$\pi$ stacking interactions of aromatic rings. A third advantage is the fact that extremely thin electrodes can be prepared, ranging from (few-layer) graphene to carbon nanotubes. Compared to the more bulky metallic electrodes, the thin carbon-based electrodes reduce the screening of an applied gate-field and therefore enhance the gate coupling.

Motivated by these advantages, carbon nanotube based nanogap electrodes have previously been constructed by oxygen-plasma etching where the gap is defined by a PMMA mask \cite{guo2006, guo2006a, whalley2007} or through electrical breakdown. \cite{Qi2004, Wei2008, Marquardt2010} To date, however, control over the gap-size below 10 nm has not been demonstrated, making it difficult to contact single molecules. Other approaches that could potentially lead to nanogap electrodes include AFM nanolithography of graphene\cite{He2010}, anisotropic etching catalyzed by nanoparticles \cite{Campos2009, Schaffel2009}, graphene nanogaps formed by mechanical stress\cite{Wang2010}, or through electrical breakdown of graphene.\cite{Standley2008}

\begin{figure} [h]
    \begin{center}
        \includegraphics[width=15cm]{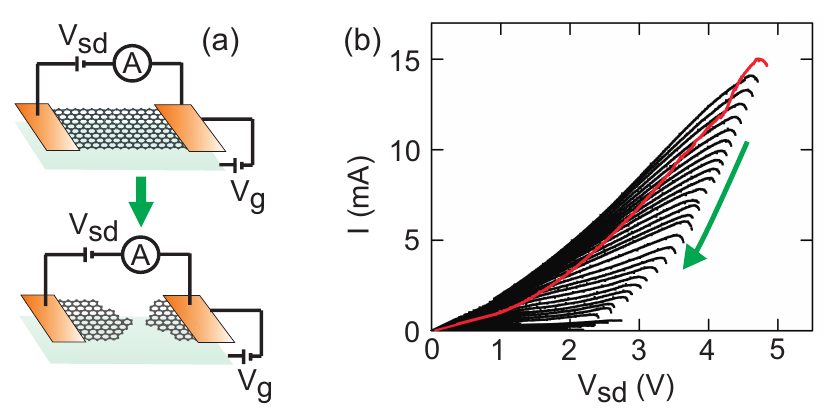}
    \end{center}
    \caption{\label{fig:design} \textbf{a)} Schematic of the feedback-controlled electroburning process, before (top) and after (bottom), the formation of nanometer sized gaps in few-layer graphite flakes. \textbf{b)} Current-voltage (I-V) traces of the evolution (green arrow) of the feedback-controlled electroburning. The first I-V trace is displayed in red.}
\end{figure}

Here, we report on the formation of nanometer-separated few-layer graphene electrodes using feedback-controlled electroburning. The process of electroburning is related to the chemical reaction of carbon atoms with oxygen at high temperatures, induced by Joule heating at high current densities. This technique has also been utilized to controllably remove shells of multi-walled carbon nanotubes \cite{Collins2001a, Collins2001, Barreiro2008}, to form nanogaps in single-walled carbon nanotubes (SWNTs) \cite{Wei2008, Qi2004, Marquardt2010} and to fabricate narrow graphene constrictions and quantum dots.\cite{Moser2009, Shi2011} An important motivation for our choice for few-layer graphene (as opposed to single layer graphene or carbon nanotubes) is that it is thin, yet its conductance largely gate-independent so that features of the contacted molecules will not be masked by the electrode's response to the gate.

We start by briefly describing our fabrication technique. Few-layer graphene flakes (between 3 - 18 nm thick) are deposited by mechanical exfoliation of kish graphite (Toshiba Ceramics) on degenerately doped silicon substrates coated with 280 nm of thermal silicon oxide. We use standard wafer protection tape as it leaves little adhesive residue on substrates. Cr/Au electrodes are patterned on top of selected few-layer graphene flakes by electron-beam lithography and subsequent metal evaporation, followed by a lift-off in cold acetone and dicholoroethane. Figure 1a (top) shows a schematic of the few-layer graphene device used for electro-burning and nanogap formation. Initial device resistances at low bias are in the order of 200 $\Omega$ - 3 k$\Omega$.

The feedback controlled electroburning is performed in air at room temperature.  The feedback control scheme is based on similar methods used for electromigration of metallic nanowires.\cite{Strachan2005, Prins2009} Typically, a voltage (V) ramp is applied to the graphite flake (1 V/s), while the current (I) is continuously recorded with a 200 $\mu$s sampling rate. The variations in the conductance (G = I/V) are monitored, with a feedback condition set at a >10 \% drop in G within the past 200 mV of the ramp. Upon the occurrence of such a drop, the voltage is swept back to zero in 10 milliseconds. Immediately after, a new sweep starts from zero voltage and the process is repeated, in this way gradually narrowing down the flake.

\begin{figure}[h]
    \begin{center}
        \includegraphics[width=15cm]{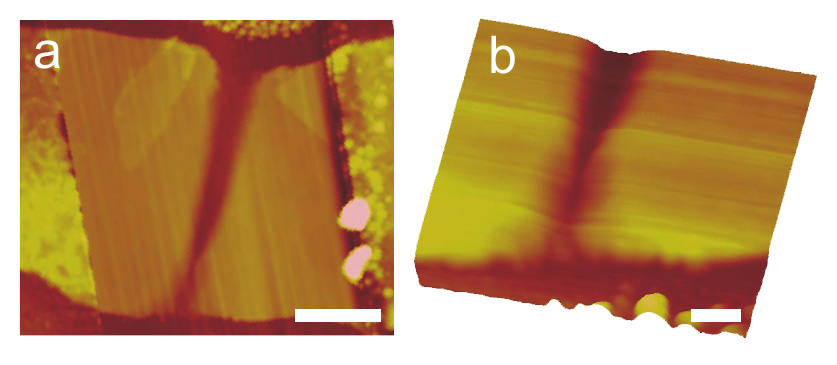}
    \end{center}
    \caption{\label{fig:molecule} \textbf{a)} AFM image of a typical graphite nanogap. The scale bar is 1 $\mu$m.  \textbf{b)} Aerial view of a zoom-in on the gap area. The scale bar is 100 nm.}
\end{figure}

\pagebreak

Figure 1b shows a typical evolution of feedback-controlled electroburning. Generally, during the first voltage ramp (red trace in figure 1b) non-linear I-V characteristics are observed, likely due to removal of contaminants on the flake by current annealing.\cite{Moser2007} Increasing the voltage further induces the first electroburning event, as can be seen from the downward curvature in the I-V characteristic, here at V = 4.8 V and I = 15 mA. The feedback then sweeps the voltage back to 0 V and a new voltage ramp is started. As the electroburning process evolves, the conductance decreases in steps and the voltage at which the electroburning occurs decreases (see green arrow in Fig. 1b). In total we have performed electroburning on 38 samples of which 35 (92~\%) underwent the electroburning process down to a low-bias resistance in the range of 500 M$\Omega$ - 10 G$\Omega$. In the other cases the feedback was not fast enough to respond, resulting in gaps with infinite  resistance (> 100 G$\Omega$).

Analyzing the cross-section of the device in figure 1b by atomic force microscopy (AFM) we calculate the critical current density at which the first electroburning event occurs, to be 5.3 x 10$^7$ A/cm$^2$. For all the devices on which we have performed the electroburning the critical current densities are comparable; between 3.8 x 10$^7$  and 7.6 x 10$^7$ A/cm$^2$, independent of the thickness of the flakes and similar to the current densities of 10$^8$ A/cm$^2$ at which single layer graphene breaks down.\cite{Barreiro2009}

To characterize the gap geometry, we have performed AFM on several devices after electroburning, a representative example of which is shown in Fig. 2. This graphite flake has a height of 12 nm, corresponding to ca. 35 layers of graphene. The image suggests that the electroburning starts from the edges in the central region of the flake, predominantly at one side. Interestingly, the height of the few-layer graphene electrodes does not change during the process of electro-burning. The evolution of the thinning can be understood by considering that the electroburning is a temperature activated process, relying on the reaction of carbon atoms in the lattice with oxygen. The highest temperature in the flake as a result of Joule heating at large current densities is reached in the central region since heat is evacuated mainly to the Au-leads, while the edge-carbon atoms are the most reactive sites due to the incomplete sp$^2$-hybridization.\cite{Warner2009, Jin2009} After the first carbon atoms have been removed on the site of highest reactivity, the electroburning will likely propagate from there as the current density and therefore the temperature is the highest near this point.

Because it is difficult to obtain an accurate gapsize from AFM characterization, we turn to the electrical characteristics of the nanogaps. Current-voltage characteristics between +/- 500 mV of 34 electro-burned samples with finite resistance were recorded at room temperature in a vacuum probe station (see Fig. 3 for an example). The junctions show current-voltage curves indicative of tunneling behavior through a single barrier. The fact that we observe tunnel currents at these low biases shows that the gaps are in the order of a few nanometers. The Simmons model can be used to estimate the gap-size\cite{Simmons1963} using  the gap-size, the barrier height and the asymmetry in the bias-voltage response as fit parameters (see supporting information for the implementation). Fits of the IV-characteristics to this model yield typical gap sizes of approximately 1-2 nm.\cite{comment3} The fitted barrier heights (< 1 eV) are lower than one would typically expect for bulk graphite. However, low barrier heights have also been observed for nanometer sized Au electrodes.\cite{Mangin2009}

\begin{figure} [h]
    \begin{center}
        \includegraphics[width=10cm]{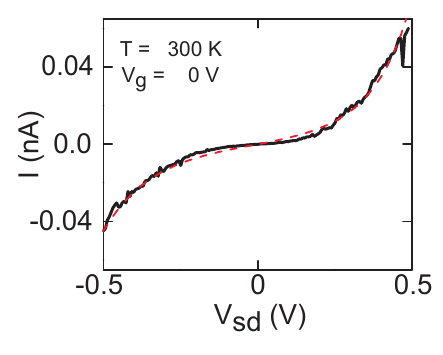}
    \end{center}
    \caption{\label{fig:potential} Representative current-voltage characteristic of a graphite nanogap (black solid line) with a fit to the Simmons model for tunneling (red dashed line). Fit parameters (accuracy of 5\%): gapsize d: 2.1 nm, prefactor A: 0.34 A eV$^{-2}$, barrier height $\phi$: 0.92 eV, asymmetry $\alpha$: -0.35 (see supporting information for more details and the definition of the parameters).}
\end{figure}

Our few-layer graphene nanogap electrodes are remarkably stable and display only small variations in the tunneling characteristics after several weeks when stored in vacuum (current-levels stay within 10\% variation)\cite{comment2}. We have also carefully measured the conductance as a function of the backgate voltage (V$_g$) at low temperature (10 K) at low bias voltages of typically 100 and 200 mV. Generally, for these devices the conductance does not vary within our experimentally accessible range of V$_g$ between +/- 40 V.\cite{comment} The small electrode separations and long-term stability of the nanogaps, combined with the absence of gate-dependent transport across the gap imply that they can be used to contact small molecules and measure three-terminal transport.

To demonstrate this, we have deposited anthracene-functionalized curcuminoid molecules (1,7-(di-9-anthracene)-1,6-heptadiene-3,5-dione, abbreviated as 9Accm, see Fig. 4a) on the nanogap devices.\cite{aliaga-alcalde2010} The anthracene-groups are extended $\pi$-conjugated systems that interact strongly with the $\pi$-system of the top graphene layer, providing a strong anchor to the electrodes, while the curcuminoid wire has a high $\pi$-electron density which can mediate charge transport. We deposit the molecules by placing the devices overnight in a chloroform solution containing 0.1 mM of 9Accm. After taking the devices out of the solution they are blow-dried by a flow of N$_2$. AFM characterization of the deposition on a reference sample shows that a sub-monolayer of molecules is formed on the devices (see supporting information). The devices are then electrically characterized in a vacuum probe station (see fig. 4b for a schematic representation).

\begin{figure*}
    \begin{center}
        \includegraphics[width=15cm]{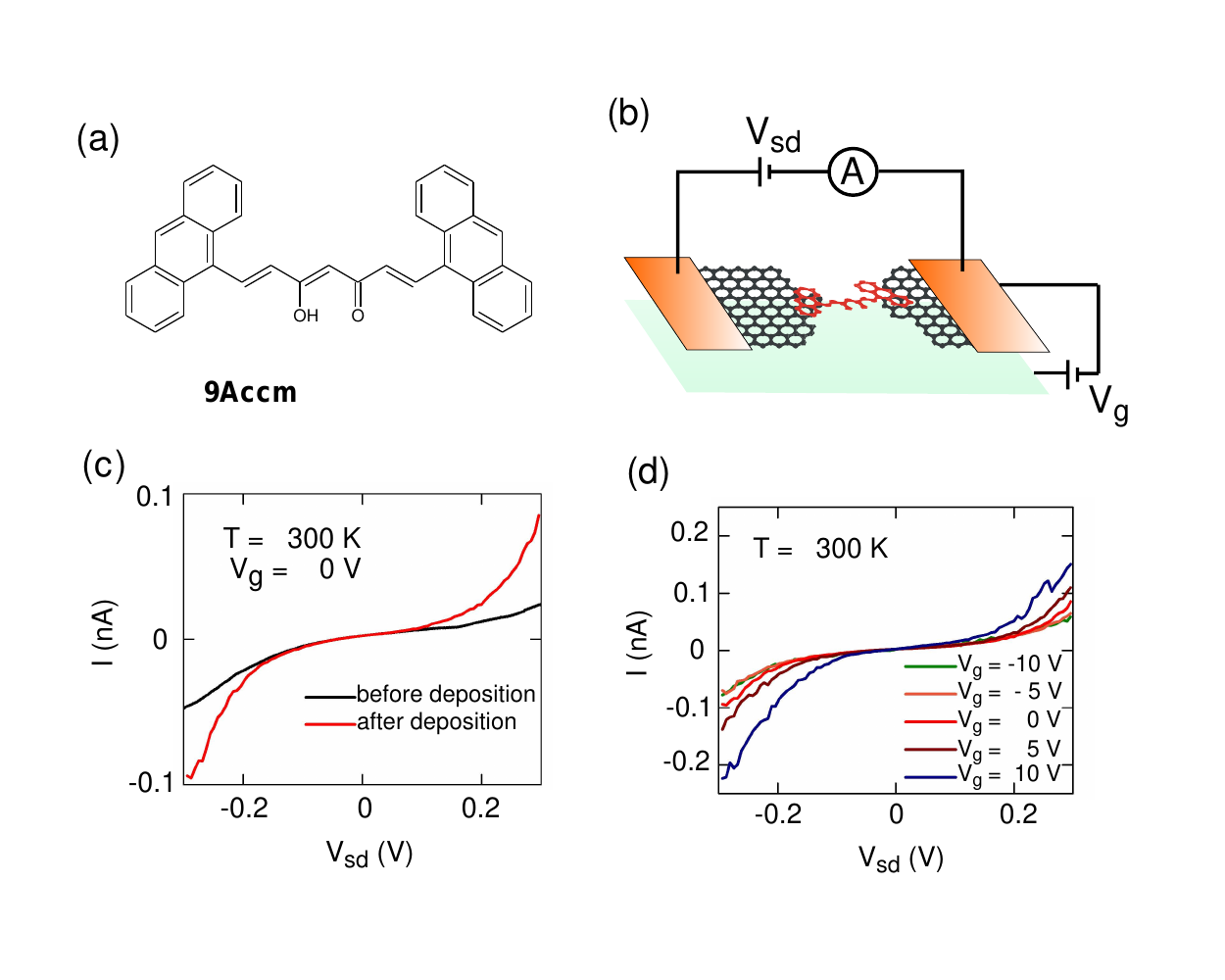}
    \end{center}
    \caption{\label{fig:dynamics} (a) Chemical structure of the anthracene terminated curcuminoid wires (1,7-(di-9-anthracene)-1,6-heptadiene-3,5-dione) (b) Artist's impression of a single 9Accm molecule bridging a graphene nanogap, representing the ideal transistor configuration. Note that the ambient conditions to which the devices are exposed may result in a thin layer of adsorbates such as water to be present on the surface of the few-layer graphene. (c) IV-characteristics of the nanogapped electrodes before and after being bridged by the 9Accm molecules at 300K. While the conductance at low bias superimposes with the empty gap characteristic, at higher bias a clear current increase is observed. (d) Dependence of the I-V characteristics of the nanogapped electrodes bridged by 9Accm molecules on the applied back-gate voltage measured at 300K.}
\end{figure*}

An important advantage of stable nanogap electrodes is that the current-voltage characteristics after deposition can be compared with the characteristics before deposition. Changes in the transport-characteristics can then be attributed to the presence of molecules in the gap. In our case, 14 out of 35 devices displayed an increase in conductance after deposition.  Figure 4c shows a typical device in which such changes in the IV-characteristic are observed. While the conductance at low bias superimposes with the empty gap characteristic, at higher bias a clear current increase is observed. Exposure of the devices to pure solvent (chloroform) does not show any significant changes in the electrical characteristics (see supporting information).

Importantly, the conductance in this device is dependent on the gate-voltage at room temperature, as illustrated in Fig. 4d. Taking current-voltage characteristics at different gate voltages between - 10 and 10 V,  the conductance increases towards more positive gate values; i.e. the blue curve in Fig. 4d displays the highest currents. The gate-dependent characteristics are robust, showing only minor variations in the conductance (<10 \%) for periods of several weeks when stored in vacuum, and even after thermal cycling to low temperatures (10 K, see below). In total we observed gate-modulated transport in 4 out of the 14 junctions that displayed an increase in conductance after deposition.

\begin{figure}
    \begin{center}
        \includegraphics[width=15cm]{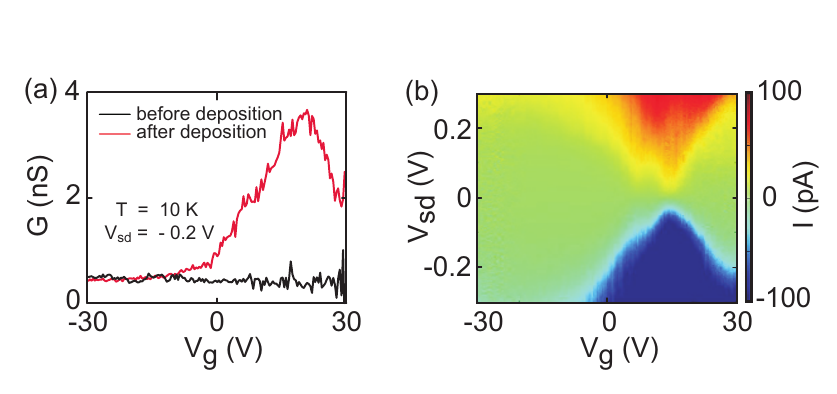}
    \end{center}
    \caption{\label{fig:conductance} (a) Conductance as a function of the applied back-gate voltage of the nanogapped electrodes bridged by 9Accm molecules at 10 K. While the empty nanogap electrodes show no dependence of the applied back-gate voltage, a clear conductance modulation as a function of V$_g$ is observed after deposition of 9Accm molecules.
(b) Current map at 10 K. IV's are taken between V$_s$$_d$ = +/- 300 mV while the backgate voltage is swept between -/+ 30 V at 100 mV intervals. In the green regions, transport is blocked due to charge quantization in the molecule, while in the red and blue regions the blockade is lifted and single-electron tunneling occurs. Although the signatures may originate from a few molecules in parallel, the single electron tunneling nature of the transport is apparent from the current map.}
\end{figure}

At low temperature (10 K), the gate-dependent transport becomes more apparent and we can moreover compare it to the empty-gap gate-dependence taken before deposition. In figure 5a the conductance at V$_s$$_d$ = - 200 mV is plotted as a function of gate voltage before and after deposition (same device as Fig. 4c and d). While the conductance is gate-independent before deposition, after deposition a clear modulation of the conductance is present towards more positive values of V$_g$, consistent with the room temperature IV-characteristics. For a full characterization, we construct conductance maps at 10 K, in which IV's are taken between V$_s$$_d$ = +/- 300 mV while the backgate voltage is swept between -/+ 30 V at 100 mV intervals. An example is shown in Fig. 5b. In the green regions, transport is blocked due to charge quantization in the molecule, while in the red and blue regions the blockade is lifted and single electron tunneling occurs. Although the signatures may originate from a few molecules in parallel, Coulomb-blockaded transport and the single-electron tunneling nature of the transport is apparent from the conductance map.

In conclusion, we report on a new method to controllably form nanogaps in few-layer graphene with nanometer separations based on feedback controlled electroburning of few-layer graphene. Gateable transport through molecules contacted between the electrodes demonstrates the potential of room-temperature operation of molecular devices. Combined with the observed stability in time, our study shows that few-layer graphene nanogaps are an interesting alternative to metal electrodes. We further note that the fabrication technique is not limited to the use of exfoliated graphene but could also be applied to CVD-grown few-layer graphene over large areas, paving the path to more complex, integrated devices involving multiple molecular devices integrated on the same chip.

\paragraph*{Acknowledgment.} \emph{We gratefully acknowledge discussions with J. M. Thijssen, and we thank A. M. Goossens for experimental help. Financial support was obtained from the Dutch Foundation for Fundamental Research on Matter (FOM), and from the EU FP7 program under the grant agreement ``ELFOS.'' N.A.A. thanks the Ministerio de Educación y Ciencia (CTQ2009-06959/BQU) and ICREA (Institució Catalana de Recerca i Estudis Avançats) for the financial support.}

\makeatletter
\renewcommand{\thefigure}{S\@arabic\c@figure}
\renewcommand{\thetable}{S\@arabic\c@table}
\makeatother

\setcounter{figure}{0}

\section{Simmons model}

\begin{figure}
    \begin{center}
        \includegraphics{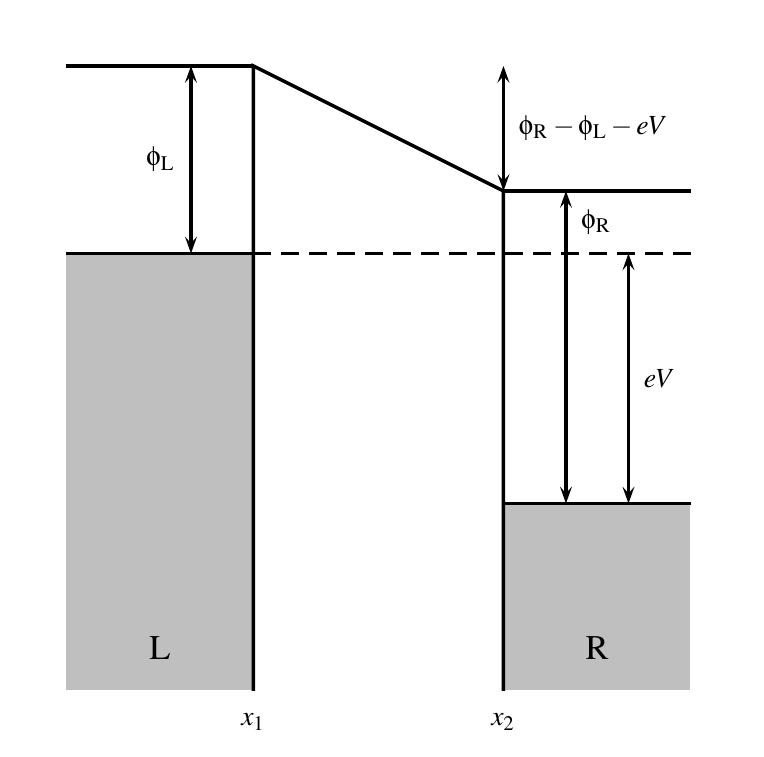}
    \end{center}
    \caption{\label{fig:barrier} Asymmetric tunnel junction with different work functions on the left ($\phi_\text{L}$) and right ($\phi_\text{R}$) leads. The barrier potential is assumed to vary linearly with the distance $x$ in the gap.}
\end{figure}

The current density in a tunnel junction with a barrier in the
$x$-direction ($\phi(x)$, see \ref{fig:barrier}) is given
by\cite{Simmons1963}
\begin{equation}\label{eq:simmons}
j=e\frac{4\pi
m}{h^3}\int_0^\infty\text{d}\epsilon\left[f_\text{L}(\epsilon)-f_\text{R}(\epsilon)\right]\int_0^\epsilon\text{d}\epsilon_xT\left(\epsilon_x\right),
\end{equation}
where
\begin{equation}
f(\epsilon)=\frac{1}{e^{-\frac{\epsilon-\mu}{k_\text{B}T}}+1}
\end{equation}
is the Fermi distribution on leads with chemical potential $\mu$,
and $T(\epsilon_x)$ is the tunnel probability of an electron with
a kinetic energy $\epsilon_x$ in the $x$-direction. In the
low-temperature limit, \emph{i.e.},
$k_\text{B}T\ll\mu_\text{L},\mu_\text{R},\phi(x)$, the Fermi
distribution effectively becomes a step-function and eq 1 reduces
to
\begin{equation}
j=e\frac{4\pi
m}{h^3}\int_{\mu_\text{R}}^{\mu_\text{L}}\text{d}\epsilon\int_0^\epsilon\text{d}\epsilon_xT\left(\epsilon_x\right).
\end{equation}
The tunnel probability can be obtained via the WKB-approximation:
\begin{equation}
T(\epsilon_x)=e^{-\frac{2}{\hbar}\int_{x_1}^{x_2}\text{d}x\left|p_x\right|}=e^{-\beta\int_{x_1}^{x_2}\text{d}x\sqrt{\phi(x)-\epsilon_x}},
\end{equation}
where $\beta=2\frac{\sqrt{2m}}{\hbar}$. For a barrier with
constant height, \emph{i.e.}, $\phi$ is independent of $x$ between
$x_1$ and $x_2$,
\begin{equation}
T(\epsilon_x)=e^{-\beta d\sqrt{\phi-\epsilon_x}},
\end{equation}
where $d=x_2-x_1$, and the current density
becomes\cite{Simmons1963}
\begin{equation}
j\approx\frac{e}{2\pi
hd^2}\left[\left(\phi-\mu_\text{L}\right)e^{-2d\frac{\sqrt{2m\left(\phi-\mu_\text{L}\right)}}{\hbar}}-\left(\phi-\mu_\text{R}\right)e^{-2d\frac{\sqrt{2m\left(\phi-\mu_\text{R}\right)}}{\hbar}}\right].
\end{equation}

This is a good approximation when the work functions of the left
and right lead are the same ($\phi_\text{L}=\phi_\text{R}$). When
they are different, the barrier height (to first order) changes
linearly from $\phi_L$ at $x_1$ to $\phi_R$ at $x_2$:
\begin{equation}
\phi(x)=\phi_\text{L}+\left(\phi_\text{R}-\phi_\text{L}\right)\frac{x-x_1}{x_2-x_1},
\end{equation}
and
\begin{equation}
\int_{x_1}^{x_2}\text{d}x\sqrt{\phi(x)-\epsilon_x}=d\frac{2}{3}\frac{\left(\phi_\text{R}-\epsilon_x\right)^{\tfrac{3}{2}}-\left(\phi_\text{L}-\epsilon_x\right)^{\tfrac{3}{2}}}{\phi_\text{R}-\phi_\text{L}}.
\end{equation}
However, this is only valid when
$\epsilon_x<\phi_\text{L},\phi_\text{R}$. When
$\phi_\text{L}<\epsilon_x<\phi_\text{R}$,
\begin{equation}
\int_{x_1}^{x_2}\text{d}x\sqrt{\phi(x)-\epsilon_x}=d\frac{2}{3}\frac{\left(\phi_\text{R}-\epsilon_x\right)^{\tfrac{3}{2}}}{\phi_\text{R}-\phi_\text{L}},
\end{equation}
and similarly for $\phi_\text{R}<\epsilon_x<\phi_\text{L}$.

For this barrier, the current density can no longer be calculated
analytically, but is easily evaluated numerically. For fitting
purposes it is generally more convenient to work with an average
barrier height $\phi=\frac{\phi_L+\phi_R}{2}$ and an asymmetry
factor $\alpha=\frac{\phi_L-\phi_R}{\phi_L+\phi_R}$. Apart from
$\phi$ and $\alpha$, also the gap size ($d$) is fitted in the
paper. The Simmons model gives the current density,  but since the
current is measured, in principle also the cross-section of the
junction would need to be fitted. However, this cross-section
cannot be fitted independently from other parameters such as the
effective electron mass in the electrodes, and it is treated as a
prefactor. Using the asymmetric Simmons model we can estimate the
gap size within approximately 5\% accuracy.

\section{Temperature dependence of the tunneling characteristics of the bare gaps}

To further investigate the nature of the electron transport in the
empty devices, we recorded IV-characteristics at different
temperatures (see Fig. S2 for a representative example, same
device as Fig. 3). In the IV-characteristics of the empty devices,
only a small decrease in the current is observed at lower
temperatures, while the characteristic shape is maintained. The
weak temperature dependence is consistent with tunneling as the
main conduction-mechanism.\cite{Simmons1963}

\begin{figure}
    \begin{center}
        \includegraphics[width=10cm]{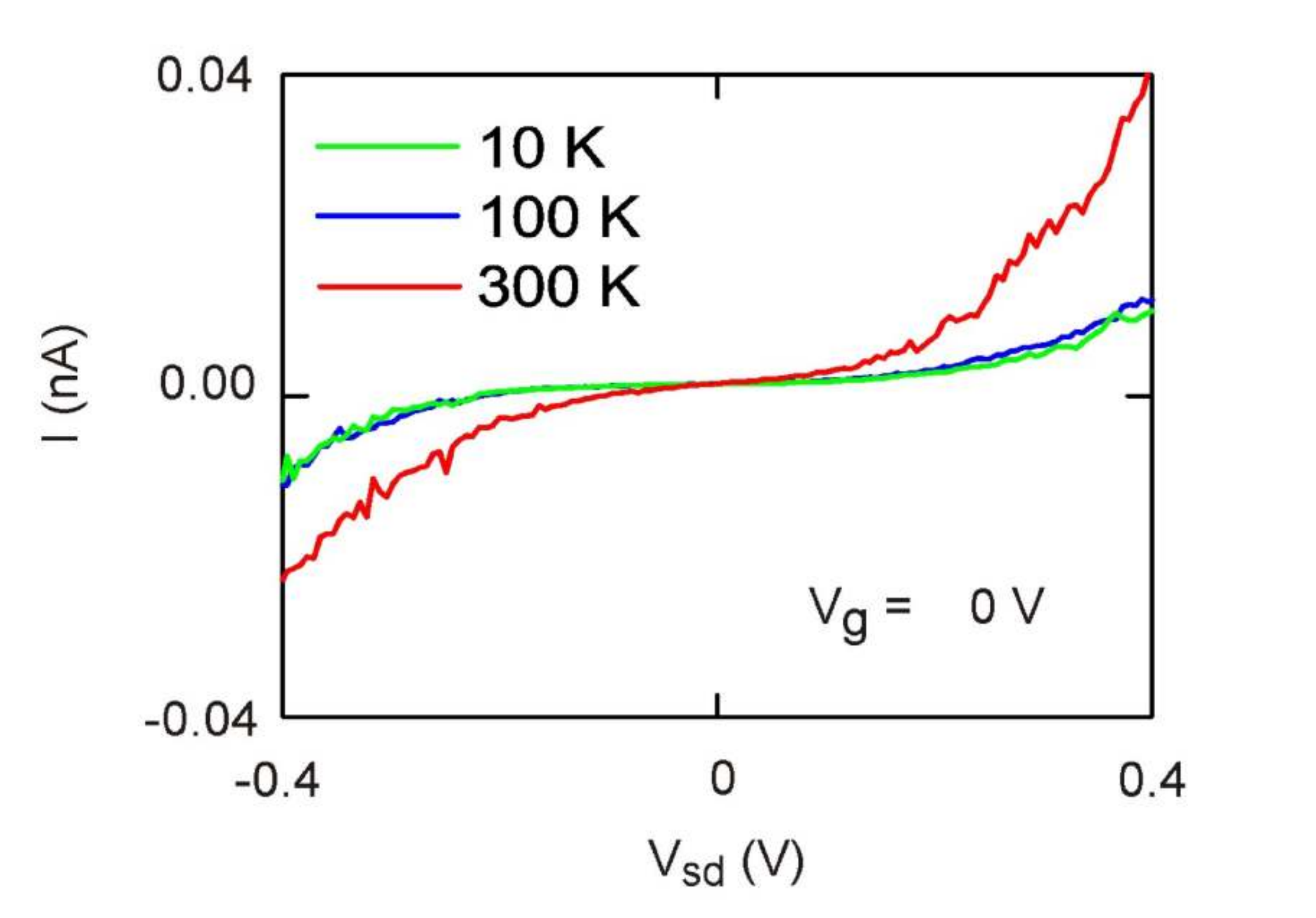}
    \end{center}
    \caption{\label{fig:barrie} Current-voltage characteristics of an empty device (same as Fig. 3) at different temperatures.}
\end{figure}

\section{Characterization of molecule deposition}

\begin{figure}
    \begin{center}
        \includegraphics[width=15cm]{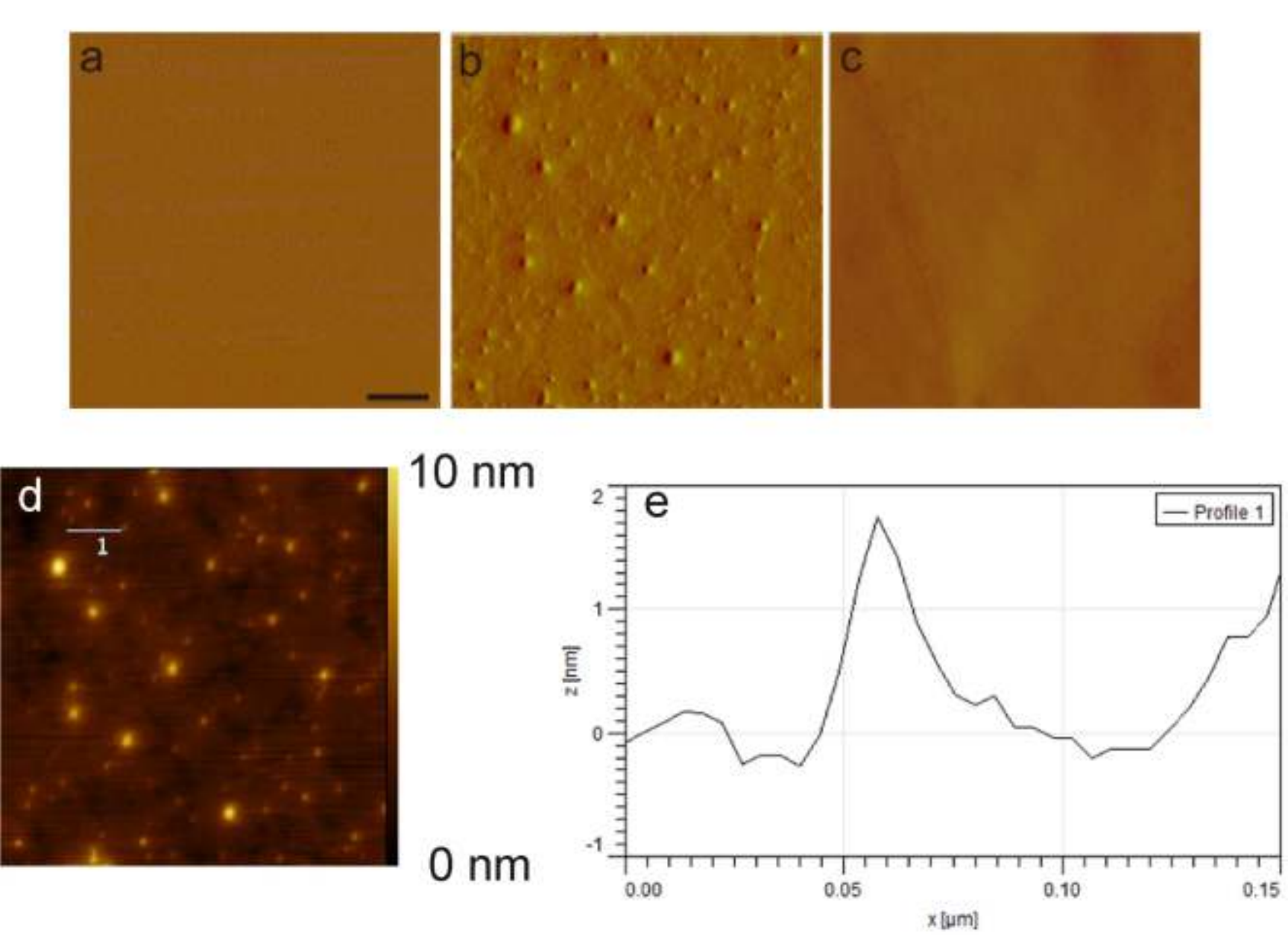}
    \end{center}
    \caption{\label{fig:barri} \textbf{a)} AFM characterization (amplitude image) of a few-layer graphene reference sample before deposition. Scale bar is 200 nm. \textbf{b)} reference sample after deposition showing that a sub-monolayer of 9Accm molecules is formed on the few-layer graphene. \textbf{c)} reference sample after exposure to only chloroform. \textbf{d)} height image of the sample after deposition (same as (b)). \textbf{e)} line trace across one of the features of (d).}
\end{figure}

We have deposited 1,7-(di-9-anthracene)-1,6-heptadiene-3,5-dione
molecules on the nanogap devices. The anthracene-groups are
extended $\pi$-conjugated systems that interact strongly with the
$\pi$-system of the top-graphene layer, providing a strong anchor
to the electrodes. We deposit the molecules by placing the devices
overnight in a chloroform solution containing 0.1 mM of 9Accm.
Chloroform was chosen as solvent as it is non-coordinating and
volatile, allowing for relatively clean deposition of molecules on
the few-layer graphene without solvent-adsorbates. After taking
the devices out of the solution they are blow-dried by a flow of
N$_2$.

We have performed careful Atomic Force Microscopy analysis of the
deposition of the molecules on the few layer graphene samples. The
surface of the few layer graphene is generally completely clean if
the mechanical exfoliation has been carefully performed and there
are no residues from the Scotch tape on top (Fig. S3a). Only after
deposition of the 9Accm molecules, small features on the surface
are observed (Fig. S3b). Exposing the few layer graphene devices
to only the clean solvent (chloroform) yields atomically smooth
surfaces if the drying of the solvent by means of a flow of N$_2$
is performed carefully and quickly (Fig. S3c). Fig S3d shows a
height image of the reference sample after deposition (same as
Fig. S3b) with a line trace taken across one of the features
displayed in Fig. S3e. The height of the features is typically in
the order of 2~nm, which is reasonable when considering that the
central part of the 9Accm molecules is not flat and some water
adsorbates might add to the height of the measurement.

\begin{figure}
    \begin{center}
        \includegraphics[width=10cm]{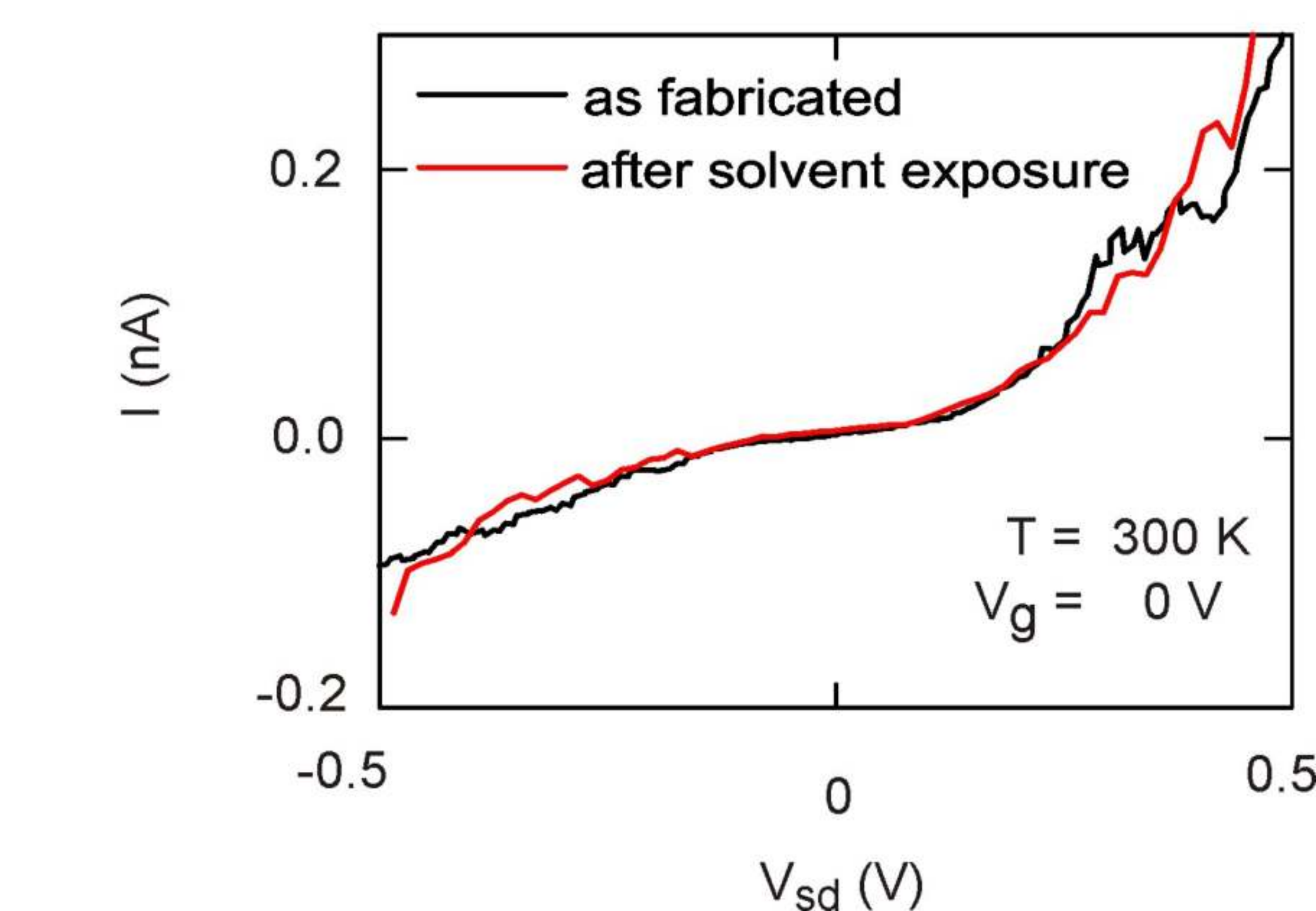}
    \end{center}
    \caption{\label{fig:barr} Current-voltage characteristics of an empty device just after fabrication (black solid line) and after exposure to pure solvent (chloroform, red solid line)}
\end{figure}

In addition, electrical control experiments of the bare nanogaps,
in which they are exposed to only pure solvent (chloroform) were
carried out. Directly after the electroburning process, the
IV-characteristics were measured. After the empty device
characterization, the samples were dipped in pure chloroform and
carefully dried underneath a flow of N$_2$. The IV-characteristics
after solvent exposure (red line in Fig. S4) display only minor
changes as compared to the devices before exposure (black line in
Fig. S4).

%\paragraph*{Supporting Information Available: Implementation of Simmons model} This material is available free of charge \emph{via} the Internet at http://pubs.acs.org.

\providecommand*{\mcitethebibliography}{\thebibliography}
\csname @ifundefined\endcsname{endmcitethebibliography}
{\let\endmcitethebibliography\endthebibliography}{}


\begin{mcitethebibliography}{38}
\bibitem[com()]{comment}
\newblock \emph{Occasionally we find a dependence of G on V$_g$, indicating
  that the sample is not fully electroburned and there still is a connection
  between the source and drain electrodes. The connection can be removed by
  performing an additional electroburning step, after which also for these
  devices we obtain few-layer graphene nanogap electrodes that do not show a
  response to the applied gate voltage.}\relax
\mciteBstWouldAddEndPunctfalse
\mciteSetBstMidEndSepPunct{\mcitedefaultmidpunct}
{}{\mcitedefaultseppunct}\relax
\EndOfBibitem
\end{mcitethebibliography}
\end{document}